\documentclass[structabstract]{aa}  
\usepackage{graphicx}
\usepackage{natbib}
\usepackage{lscape}
\usepackage{rotating}
\usepackage{txfonts}
\newcommand{\ergs}{erg\ s$^{-1}$}
\newcommand{\kms}{km\ s$^{-1}$}

\newcommand{\xmm}{{\sc XMM}\emph{-Newton}}
\newcommand{\ros}{\emph{{\sc ROSAT}}}
\newcommand{\ch}{\emph{{\sc Chandra}}}

\newcommand{\loglxlb}{$\log[L_{\rm X}/L_{\rm BOL}]$}
\begin{document}
   \title{The first X-ray survey of Galactic Luminous Blue Variables\thanks{Based on observations collected with \xmm, an ESA Science Mission with instruments and contributions directly funded by ESA Member States and the USA (NASA), and with \ch.}}

   \author{Ya\"el Naz\'e\inst{1}\fnmsep\thanks{Research Associate FRS-FNRS}
           \and Gregor Rauw \inst{1} \and Damien Hutsem\'ekers\inst{1}\fnmsep\thanks{Senior Research Associate FRS-FNRS}
          }

   \institute{GAPHE, D\'epartement AGO, Universit\'e de Li\`ege, All\'ee du 6 Ao\^ut 17, Bat. B5C, B4000-Li\`ege, Belgium\\
              \email{naze@astro.ulg.ac.be}
             }

 
  \abstract
   {}
   {The X-ray emission of massive stars has been studied when these objects are in their main-sequence phase, as well as in their Wolf-Rayet phase. However, the X-ray properties of the transitional Luminous Blue Variable (LBV) phase remain unknown.}
   {Using a dedicated but limited \xmm\ survey as well as archival \xmm\ and \ch\ observations, we performed the first X-ray survey of LBVs: about half of the known LBVs or candidate LBVs are studied.}
   {Apart from the well known X-ray sources eta Car and Cyg\,OB2\,\#12, four additional LBVs are detected in this survey, though some doubt remains on the association with the X-ray source for two of these. For the other LBVs, upper limits on the flux were derived, down to $\log[L_{\rm X}/L_{\rm BOL}]-9.4$ for P\,Cyg. This variety in the strength of the X-ray emission is discussed, with particular emphasis on the potential influence of binarity. }
   {}


   \maketitle
%

\section{Introduction}
Luminous Blue Variables (LBVs), also called S-Doradus variables, form 
a class of peculiar massive stars mostly found in the top-left part of the Hertzsprung-Russell
diagram (for a review, see \citealt{hum94}). Their main characteristics 
are a high mass-loss rate (up to 10$^{-4}$\,$M_{\odot}$\,yr$^{-1}$), a high 
luminosity ($\sim10^6$\,$L_{\odot}$), and significant spectroscopic 
as well as photometric variability. These variations occur on several 
timescales and amplitudes, from common small-scale 
changes of about 0.1mag to rare giant eruptions with 2.5mag variation. 

The most emblematic objects 
in this category are those which have undergone unexpected, enormous 
variations (P\,Cyg in the 17th century, eta\,Car in the 19th century, 
and HD\,5980 in 1993--5 and 1960--5). However, these dramatic cases should 
not be considered as the best representative objects of the class 
\citep{dav89}.

LBVs are often surrounded by young and massive circumstellar nebulae, 
which result from strong mass-loss episodes (see reviews by
\citealt{lam89,smi11}). 

In view of their variability and the presence of nitrogen-enriched circumstellar 
nebulae, LBVs are thought to be associated to a short ($<10^5$\,yr) 
unstable stage in the life of massive stars, between the main-sequence 
phase and the late Wolf-Rayet phase \citep[e.g.][]{lam01}. Their crucial role in the 
mass-loss process of massive stars has been revived in the last few years 
as the mass-loss rates of O and WR stars were revised downwards \citep[e.g.][]{smi08}.

While dedicated campaigns of X-ray observations have targeted massive stars 
in their O-star phase as well as in their Wolf-Rayet phase, not much was done
for studying LBVs. Only the X-ray bright sources eta\,Car \citep{sew79,cor10}, Cyg OB2 \#12
\citep{har79,rau11} and HD\,5980 \citep{naz02,naz04,naz07}, which display X-ray 
luminosities of 10$^{34..35}$\,\ergs, have received attention. The sole 
exception may be P\,Cyg, whose \ros\ HRI data are discussed in length by 
\citet{ber00} and mentioned by \citet{osk05a}. This observation led to an 
upper limit on the luminosity of $\sim10^{31}$\,\ergs. Finally, non-detections 
or a very weak detection of a few candidate LBVs are sometimes reported
when discussing the X-ray emission of their home clusters, but without 
much detail: Pistol Star \citep{mun06,osk05b}, Sher\,25 \citep{mof02} 
and W243 \citep{cla08}. 

This paper aims at bridging the gap between X-ray studies of O and WR stars
by using the most sensitive X-ray observatories available at the present 
time, \xmm\ and \ch. Using a dedicated (though limited) survey as well 
as archival data, we investigate the X-ray emission of about half 
of the known LBVs. The paper is organized as follows: Sect. 2 presents 
the LBV catalog used for our survey, Sect. 3 shows the available X-ray 
observations and their reduction, Sect. 4 lists our results, Sect. 5 
discusses them and Sect. 6 finally summarizes the new information 
and concludes.

\section{The sample}
Two main S-Dor or LBV catalogs are available in the literature: 
\citet{van01} and \citet{cla05}. The latter lists more 
sources - 12 confirmed and 23 candidates (which will be noted cLBVs 
in the following) - and includes the 21 objects from van Genderen, 
with the exception of IRC+10420 and HD148937 (which is of the peculiar Of?p 
type, see \citealt{wal72} for initial definition of this class and 
\citealt{naz11c} for an in-depth study of this object in particular).

Since Clark et al.'s work, two new cLBVs have been reported in the 
literature: MWC930 \citep{mi05} and 2MASS J17460562--2851319 (also 
known as G0.120--0.048, \citealt{mu10}). Both were added to our catalog.

Recently, Spitzer surveys of the Galactic plane (e.g. MIPSGAL) have 
led to the discovery of many ring nebulae surrounding luminous central 
stars \citep{wac10,gva10} and follow-up observations were conducted in 
order to determine the nature of these objects.
 
\citet{wac10} list 4 known LBVs (or candidates) and 6 additional 
candidate LBVs. They also have 6 Be candidates for which they mention that
their separation from the cLBVs is arbitrary and that these objects should 
therefore be considered as potential LBVs too. \citet{gva10} reported the
detection of 2 additional cLBVs and a (known, though not included in the
Clark et al. compilation) cLBV. Note that one of these, MN112, was 
identified as Be in Wachter et al.'s list (under the name 2MASS
J19443759+2419058).

Expanding on these works, \citet{wac11} list 8 additional
LBV candidates (one having been classified as Be in her previous paper) 
and 4 additional Be candidates (two already classified so in their 
previous paper). They also mention finding 6 additional LBV candidates 
and one Be candidate from the follow-up observations of the sample of 
\citet{gva10}. 

Considering these new detections, we sum up Clark's list of 35 
LBVs or cLBVs, 2 recent additions, 21 objects from Wachter's sample and 9 from 
Gvaramadze's sample. This enlarges the (c)LBV catalog to 67 objects. 
They are given in Table\,\ref{cat}, along with their main properties, 
when known (see quoted references). Typical errors on $\log(T)$
and $\log(L)$ are 0.05 and 0.2 dex, respectively \citep{van01}. More generally, it must 
be kept in mind that the properties of LBVs are much more uncertain than 
those of O and WR stars. Names and positions come from the Simbad database
and \citet{gva10} for their sample - RA and DEC were rounded to 2 decimal digits. For GRS 79.29+0.46, the Simbad and van 
Genderen coordinates do not agree
well, and a revised estimate of the position was found in \citet{jim10}: 
the coordinates quoted in Table \ref{cat} correspond to those of 
2MASS J20314228+4021591, the (bright) 2MASS source closest to the Simbad position as well as to the van Genderen and Jim\'enez-Esteban coordinates.

\begin{table*}
\caption{Catalog of the known LBVs and LBV candidates (see Sect. 2), X and C refer to \xmm\ and \ch\ observations, respectively.}
\label{cat}
\begin{center}
\begin{tabular}{lccccccccc}
            \hline\hline
Simbad Name & RA(J2000) & DEC(J2000) & $\log(T_{\rm eff})$ & $\log(L_{\rm BOL})$ & $E(B-V)$ & $d$ & Ref\tablefootmark{a} & X ? & C ? \\
 & hh mm ss & $^{\circ}$ ' '' &  K & $L_{\odot}$ & & kpc & & & \\
\hline
\multicolumn{10}{l}{\it Clark's catalog}\\
\multicolumn{10}{l}{\it A. Confirmed LBVs}\\
eta Car                 & 10 45 03.59    & --59 41 04.26    & 4.36-4.18  &  5.14-6.34  &  0.50  &    2.3  &   vG01        &x&x\\
P Cyg                   & 20 17 47.20    & +38 01 58.55     & 4.26       &  5.70       &  0.51  &    1.7  &   vG01        &x& \\
HD 168607               & 18 21 14.89    & --16 22 31.76    & 3.97       &  5.38       &  1.55  &    2.2  &   vG01        & & \\
AG Car                  & 10 56 11.58    & --60 27 12.81    & 4.46-4.13  &  6.14-6.22  &  0.63  &    6.0  &   vG01        &x& \\
HR Car                  & 10 22 53.84    & --59 37 28.38    & 4.34-3.90  &  5.90-5.62  &  1.00  &    5.2  &   vG01        & & \\
HD 160529               & 17 41 59.03    & --33 30 13.71    & 3.95       &  5.46       &  1.10  &    2.5  &   vG01        &x& \\
WRAY 15-751             & 11 08 40.06    & --60 42 51.7     & 4.48       &  5.91       &  1.60  &    6.0  &   pa06,vG01   & & \\ 
qF 362                  & 17 46 17.98    & --28 49 03.46    & 4.05       &  6.25       &  8.00  &   11.5  &   na09        &x&x\\
AFGL 2298               & 19 00 10.89    & +03 45 47.1      & 4.01-4.07  &  6.11-6.30  &  9.00  &   10.0  &   cl09b       & & \\ 
GAL 024.73+00.69        & 18 33 55.29    & --06 58 38.7     & 4.08       &  5.6        &  3.40  &    5.2  &   cl03        &x& \\ 
Cl* Westerlund 1 W 243  & 16 47 07.50    & --45 52 29.16    & 3.93       &  5.86       &  4.35  &    4.5  &   ri09,cl05   &x&x\\
GCIRS 34W               & 17 45 39.73    & --29 00 26.50    & 4.29       &  5.5        & 10.80  &    7.6  &   ma07        &x&x\\
\multicolumn{10}{l}{\it B. Candidate LBVs}\\
NAME VI Cyg 12          & 20 32 40.96    & +41 14 29.28     &  4.11      &   6.42      &  3.40  &    1.7  &   vG01        &x&x\\
GRS 79.29+0.46          & 20 31 42.28    & +40 21 59.13     &  4.40      &   6.30      &  5.00  &    2.0  &   vG01        & &x\tablefootmark{b}\\
Pistol Star             & 17 46 15.24    & --28 50 03.58    &  4.07      &   6.20      &  8.00  &   11.5  &   na09        &x&x\\
HD 168625               & 18 21 19.55    & --16 22 26.06    &  4.08      &   5.34      &  1.50  &    2.2  &   vG92        & & \\
zet01 Sco               & 16 53 59.73    & --42 21 43.31    &  4.26      &   6.10      &  0.66  &    2.0  &   cr06        & & \\
HD 326823               & 17 06 53.91    & --42 36 39.74    &  4.34      &   5.30      &  1.15  &    2.0  &   mr07        & & \\
HD 316285               & 17 48 14.04    & --28 00 53.13    &  4.18      &   5.44      &  1.81  &    1.9  &   vG01        &x& \\
Hen 3-519               & 10 53 59.59    & --60 26 44.31    &  4.48      &   6.26      &  1.30  &    8.0  &   vG01        & & \\
HD 80077                & 09 15 54.79    & --49 58 24.58    &  4.23      &   6.30      &  1.47  &    3.0  &   vG01        & & \\
AS 314                  & 18 39 26.11    & --13 50 47.19    &  4.01      &   4.90      &  0.90  &    8.0  &   vG01        & & \\
MWC 314                 & 19 21 33.97    & +14 52 56.89     &  4.48      &   6.34      &  1.84  &    3.0  &   vG01        & & \\
GRS 25.5+0.2            & 18 37 05.21    & --06 29 38.0     &            &             & 10.00  &   14.5  &   vG01        &x& \\ 
GAL 026.47+00.02        & 18 39 32.24    & --05 44 20.5     &  4.23      &   6.0       &  3.80  &    6.5  &   cl03        & &x\\
WRAY 17-96              & 17 41 35.45    & --30 06 38.8     &  4.11      &   6.26      &  2.84  &    4.5  &   eg02        & & \\ 
WR102ka                 & 17 46 18.12    & --29 01 36.6     &  4.40      &   6.5       &  8.0   &    8.0  &   ba08        &x&x\\ 
NAME LBV 1806-20        & 18 08 40.31    & --20 24 41.1     &            &   6.3       &  9.4   &   11.8  &   fi04,ei04   &x&x\\ 
NAME Sher 25 star       & 11 15 07.8     & --61 15 17       &  4.35      &   5.9       &  1.60  &    6.3  &   sm02        & &x\\  
$[$OMN2000$]$ LS1       & 19 23 47.64    & +14 36 38.4      &  4.12-4.14 &   5.75      &  3.50  &    6.0  &   cl09a       & &x\\ 
GCIRS 16NE              & 17 45 40.26    & --29 00 27.09    &  4.29      &   5.9       &  8.30  &    7.6  &   ma07        &x&x\\
GCIRS 16C               & 17 45 40.13    & --29 00 27.64    &  4.24      &   5.9       &  8.30  &    7.6  &   ma07        &x&x\\
GCIRS 16SW              & 17 45 40.12    & --29 00 29.06    &            &             &        &         &               &x&x\\
GCIRS 16NW              & 17 45 40.05    & --29 00 26.87    &            &             &        &         &               &x&x\\
GCIRS 33SE              & 17 45 40.02    & --29 00 31.0     &  4.26      &   5.75      &  9.00  &    7.6  &   ma07        &x&x\\ 
\hline
\multicolumn{10}{l}{\it New additions (cLBVs)}\\
2MASS J17460562--2851319 & 17 46 05.625 & --28 51 31.92      &            &   6.2-6.6   &  8.00  &   11.5  &   mu10        &x&x\\
MWC 930                 & 18 26 25.24  & --07 13 17.8       &  4.34      &   5.5       &  2.5   &    3.5  &   mi05        &x& \\
\hline
\multicolumn{10}{l}{\it Wachter's sources (cLBVs)}\\
2MASS J15484207--5507422 & 15 48 42.07 & --55 07 42.21 & & & & & & & \\
2MASS J16290377--4746264 & 16 29 03.78 & --47 46 26.48 & & & & & & & \\
2MASS J16364278--4656207 & 16 36 42.78 & --46 56 20.73 & & & & & & &x\tablefootmark{b}\\
2MASS J17082913--3925076 & 17 08 29.14 & --39 25 07.68 & & & & & & & \\
2MASS J17110094--3945174 & 17 11 00.94 & --39 45 17.45 & & & & & &x& \\
2MASS J18415965--0515409 & 18 41 59.65 & --05 15 40.93 & & & & & & &x\\
2MASS J16431636--4600424 & 16 43 16.37 & --46 00 42.42 & & & & & & & \\
2MASS J16493770--4535592 & 16 49 37.70 & --45 35 59.27 & & & & & & & \\
2MASS J17435981--3028384 & 17 43 59.85  & --30 28 38.5  & & & & & & & \\ 
2MASS J18133121--1856431 & 18 13 31.21 & --18 56 43.20 & & & & & & & \\
2MASS J19325284+1742303  & 19 32 52.85 & +17 42 30.33 & & & & & & & \\
2MASS J19443759+2419058\tablefootmark{c} & 19 44 37.60 & +24 19 05.87 & & & & & gv10 & & \\
2MASS J16313781--4814553 & 16 31 37.82 & --48 14 55.30 & & & & & & & \\
2MASS J16461734--4508478 & 16 46 17.35 & --45 08 47.85 & & & & & &x& \\
2MASS J17374754--3137333\tablefootmark{d} & 17 37 47.54 & --31 37 33.38 & & & & & & & \\
2MASS J17374730--3137370\tablefootmark{d} & 17 37 47.31 & --31 37 37.07 & & & & & & & \\
2MASS J17421401--2955360 & 17 42 14.02 & --29 55 36.06 & & & & & & & \\
2MASS J18022233--2238002 & 18 02 22.34 & --22 38 00.24 & & & & & &x& \\
2MASS J18455593--0308297 & 18 45 55.94 & --03 08 29.72 & & & & & & & \\
2MASS J18510295--0058242 & 18 51 02.95 & --00 58 24.21 & & & & & & & \\
2MASS J18070516--2015163 & 18 07 05.17 & --20 15 16.31 & & & & & & & \\
2MASS J19011669+0355108  & 19 01 16.69 & +03 55 10.81  & & & & & & & \\
\hline
\end{tabular}
\end{center}
\end{table*}

\setcounter{table}{0}

\begin{table*}
\caption{Continued}
\begin{center}
\begin{tabular}{lccccccccc}
            \hline\hline
Simbad Name & RA(J2000) & DEC(J2000) & $\log(T_{\rm eff})$ & $\log(L_{\rm BOL})$ & $E(B-V)$ & $d$ & Ref\tablefootmark{a} & X ? & C ? \\
 & hh mm ss & $^{\circ}$ ' '' &  K & $L_{\odot}$ & & kpc & & & \\
\hline
\multicolumn{10}{l}{\it Gvaramadze's sources (cLBVs)}\\
MN13  & 13 42 33.08 & --62 48 11.3 & & & & & & & \\
MN39  & 16 10 26.55 & --51 21 25.3 & & & & & & & \\
MN41  & 16 26 34.28 & --50 21 01.9 & & & & & & & \\
MN53  & 17 09 24.78 & --40 08 45.6 & & & & & &x&x\\
MN79  & 18 28 33.41 & --11 46 44.2 & & & & & &x&x\\
MN101 & 19 06 24.54 &  08 22 01.6 & & & & & & & \\
MN107 & 19 24 03.34 &  13 39 49.4 & & & & & & & \\
MN46  & 16 43 16.37 & --46 00 42.4 & & & & & & & \\
MN83  & 18 39 23.01 & --05 53 19.9 & & & & & & &x\\
\hline	
\end{tabular}
\end{center}
\tablefoottext{a}{References are: ba08 = \citet{ba08}, cl03 = \citet{cl03}, cl05 = \citet{cl05}, cl09a =\citet{cl09b}, cl09b =\citet{cl09a}, cr06 = \citet{cr06}, eg02 = \citet{eg02}, ei04 = \citet{ei04}, fi04 = \citet{fi04}, gv10 = \citet{gv10}, ma07 = \citet{ma07}, mr07 = \citet{mr07}, mu10 = \citet{mu10}, mi05 = \citet{mi05}, na09 = \citet{na09}, pa06 = \citet{pa06}, ri09 = \citet{ri09}, sm02 = \citet{sm02}, vG92 = \citet{vG92}, vG01 = \citet{van01}.
}\\   
\tablefoottext{b}{Not yet publicly available.}\\
\tablefoottext{c}{In \citet{gva10}, this source appears as MN112.}\\
\tablefoottext{d}{\citet{wac11} quote as LBV candidate their source \#27 of \citet{wac10}. However, in the latter paper, there are two \#27 sources, a and b. Both are thus quoted here together.}
\end{table*}

\section{X-ray observations}
We have obtained XMM observations for a limited LBV survey 
(ObsID=0600030, PI Naz\'e). To enlarge the sample as much as 
possible, we further searched the archives for additional, serendipitous 
observations of the known LBVs or cLBVs lying within 15' (for \xmm) or 
10' (for \ch) of the on-axis position. Only imaging exposures 
 (i.e. no XMM ``timing'' mode or Chandra ``continuous clocking'' mode) longer than 5ks were 
considered. Indeed, the quality of these serendipitous observations 
is often not as good as for the requested on-axis data, as the source usually 
lies far off-axis and the effective exposure time may not be very long. 

In total, including our targets, \xmm\ has observed 26 (c)LBVs while 
\ch\ has observed 21 objects. A closer look indicates that 16 targets 
are in common, implying that 31 (c)LBVs, or half of the catalog, have 
been observed in X-rays (see crosses in the last columns of Table \ref{cat}). Note that, 
if we restrict ourselves to the Clark catalog, \xmm\ has observed 19 objects and
\ch\ 16, with 13 sources in common: X-ray observations therefore exist 
for two-thirds of Clark's catalog. 

Two final remarks must be made. First, the datasets for the two well-known X-ray 
emitters eta\,Car and Cyg OB2 \#12 have not been re-reduced or re-analyzed, 
as many published papers already discuss them in detail (see Sects. 4.1.1 and 4.1.3).
Second, the analysis of the ten sources close to the Galactic center relies
on the \ch\ data only, as \xmm\ is not very good at disentangling sources 
in this crowded region. 

\subsection{\xmm}
The \xmm\ EPIC data were reduced in a standard way using SAS v10.0. After 
the pipeline processing (tasks {\it emproc, epproc}), the recommended pattern 
filtering was applied (${\rm PATTERN}\le12$ and XMMEA\_EM filter for MOS, 
${\rm PATTERN}\le4$ and ${\rm FLAG}=0$ for pn). The event files were further filtered 
to reject times affected by background flares. Images in the 0.5--8.0\,keV 
domain were then produced and used for the detection algorithm {\it 
edetectchain} when only one exposure was available\footnote{See the analysis 
threads for detail on its use: http://xmm.esac.esa.int/sas/current/documentation/threads/ }. 
The logarithmic likelihood for detection was put at 10, equivalent to a 
4$\sigma$ detection. Sensitivity maps for likelihoods $\log L=2.3$ (to be consistent 
with the Galactic Center data, see Sect. 3.4) and combining all instruments 
were also produced using the task {\it esensmap}. These maps provide 
equivalent on-axis limiting count rates, which were transformed into 
actual fluxes using WebPimms\footnote{http://heasarc.gsfc.nasa.gov/Tools/w3pimms.html}. 
The chosen model is a simple optically-thin
thermal emission with a temperature of 0.6\,keV and an absorption directly related
to the known extinctions of the targets (using 
$N_{\rm H}=5.8\times 10^{21}\times E(B-V)$\,cm$^{-2}$, \citealt{boh78}). Using 
temperatures of 0.3 or 1.0\,keV does not change the results by much: for an 
observed flux of 10$^{-15}$\,erg\,cm$^{-2}$\,s$^{-1}$ and a representative absorbing 
column of 10$^{22}$\,cm$^{-2}$, the total EPIC count rates vary by 10\%
in these cases compared to the 0.6\,keV case. As our limits are certainly 
not more accurate than 10\%, it means that they are quite insensitive to the adopted
temperature. For the Wachter et al. or Gvaramadze et al. sources,
no information on distance, bolometric luminosity or reddening is
available. To get a limit in (observed) flux, we assumed a representative absorbing column 
of 10$^{22}$\,cm$^{-2}$. Changing the latter by an order of magnitude (i.e. 
10$^{21}$ or 10$^{23}$\,cm$^{-2}$) would change the total EPIC 
count rate corresponding to a given observed flux, whatever the filter, by 
a factor of $\sim2$. We therefore consider that the flux limits for 
these objects are accurate within a factor of 2.

For \ch\ data, the upper limit determination is slightly different (see next section),
but we checked on the \xmm\ data of AG Car that both methods provide
similar estimates, and thus are fully compatible with each other.

When several exposures of a given region exist, a recent task added to the 
SAS, {\it emosaicproc}, should help combine them. However, its use is not 
straightforward. For 2MASS J18022233--2238002, two of the three exposures 
had a strange bug in the attitude file which prevents their use with this 
task (see also below, in the discussion of this source). For W243, the 
start time of the Rev 1499 summary file was wrong and had to be corrected 
by hand before the task could work. In addition, while the detection 
algorithm ran otherwise smoothly, the creation of a sensitivity map 
combining all exposures and all instruments is not always possible, 
as the task {\it esensmap} only accepts combinations of $<10$ input files. 
However, the effective area is not always the dominant factor limiting
the sensitivity of the \xmm\ data (see Sects. 4.1.2 and 4.3.10) and the impact
of this problem is thus quite limited.

Table \ref{xmm} lists the results of our analysis of the \xmm\ data. 

Finally, a check was made using the \xmm\ upper limit 
server\footnote{http://xmm.esac.esa.int/UpperLimitsServer/}. In 3 cases, 
only slew data were used, leading to very large upper limits compared to 
data from pointed observations, as could be expected. In addition, 
several detections with varying count rates were reported for W243 as well 
as a detection for P\,Cyg, but they are due to source confusion in these 
regions (see details on these sources below). In all cases apart from those just discussed, the server 
finds upper limits which are compatible with ours, though slightly higher.

\begin{table*}
\caption{\xmm\ imaging observations of (c)LBVs. A null off-axis angle (OFA) indicates an on-axis observation.}
\label{xmm}
\begin{center}
\begin{tabular}{lccccccc}
            \hline\hline
Simbad Name & Obsid & Rev & OFA & EPIC count rate & $F^{\rm obs}_{\rm X}$ & $L^{\rm unabs}_{\rm X}$ & \loglxlb \\
 & & & (\arcmin)& ct\,s$^{-1}$ & erg\,cm$^{-2}$\,s$^{-1}$ & \ergs & \\
\hline
P Cyg                   & 0600030201 &1891 & 0. & $<$0.0010 & $<1.0\times 10^{-15}$ & $<8.3\times 10^{29}$ & $<-9.4$\\
AG Car                  & 0600030101 &1853 & 0. & $<$0.0009 & $<8.1\times 10^{-16}$ & $<1.0\times 10^{31}$ & $<-8.7..8.8$ \\
HD 160529               & 0600030701 &1797 & 0. & $<$0.0008 & $<7.8\times 10^{-16}$ & $<3.2\times 10^{30}$ & $<-8.5$\\
GAL 024.73+00.69        & 0301880301 &1145 & 5.6& $<$0.006  & $<8.4\times 10^{-15}$ & $<7.2\times 10^{32}$ & $<-6.3$\\ 
Cl* Westerlund 1 W 243  & 0410580601 &1317 & 0.5& $<$0.0017 & $<2.6\times 10^{-15}$ & $<2.3\times 10^{32}$ & $<-7.1$\\ 
                        & 0505290201 &1409 & 0.5& & & & \\ 
                        & 0505290301 &1499 & 0.5& & & & \\ 
                        & 0555350101 &1593 & 0.5& & & & \\ 
                        & 0604380101 &1778 & 0.5& & & & \\
HD 316285               & 0112970101 &0145 & 7.8& $<$0.0013 & $<1.4\times 10^{-15}$ & $<6.1\times 10^{30}$ & $<-8.2$\\
GRS 25.5+0.2            & 0400910301 &1256 & 10.& $<$0.0025 & $<5.7\times 10^{-15}$ & $<1.7\times 10^{34}$ & \\ 
NAME LBV 1806-20\tablefootmark{a}        & 0205350101 &0869 & 0. & $<$0.01   & $<2.2\times 10^{-14}$ & $<4.1\times 10^{34}$ & $<-5.3$\\ 
                        & 0164561101 &0884 & 0. & & & & \\ 
                        & 0164561401 &1066 & 0. & & & & \\ 
                        & 0406600301 &1157 & 0. & & & & \\ 
                        & 0406600401 &1237 & 0. & & & & \\ 
                        & 0502170301 &1428 & 0. & & & & \\ 
                        & 0502170401 &1523 & 0. & & & & \\ 
                        & 0554600301 &1601 & 0. & & & & \\ 
                        & 0554600401 &1691 & 0. & & & & \\ 
MCW 930\tablefootmark{b}& 0650591501 &2059 & 7.5& $<$0.0006 & $<1.9\times 10^{-15}$ & $<5.0\times 10^{31}$ & $<-7.4$\\
2MASS J17110094--3945174 & 0502080301 &1431 & 0.9& $<$0.003  & $<3.2\times 10^{-15}$ & & \\
2MASS J16461734--4508478 & 0164561001 &0873 & 2.6& $<$0.004  & $<4.9\times 10^{-15}$ & & \\
2MASS J18022233--2238002 & 0135742801 &0600 & 11.& $<$0.0035 & $<3.8\times 10^{-15}$ & & \\
MN53\tablefootmark{b}   & 0148690101 &0681 & 6.9& $<$0.0005 & $<1.5\times 10^{-15}$ & & \\
MN79                    & 0051940401 &0229 & 3.7& $<$0.0012 & $<1.3\times 10^{-15}$ & & \\
\hline
eta Car                 & \multicolumn{4}{c}{see \citet{cor10}} & $0.06..3\times 10^{-10}$ & & $\sim$--5\\
NAME VI Cyg 12          & \multicolumn{4}{c}{see \citet{rau11}} & $2.7\times 10^{-12}$ & $8.2\times 10^{33}$ & $-6.1$\\
\hline	
\end{tabular}
\end{center}
\tablefoot{Column 
1 provides the (c)LBV name, columns 2 to 4 the details of the observation(s) 
used, and the next columns list the derived EPIC count rates, fluxes, and 
luminosities in the 0.5--8.0\,keV energy range. Limits correspond to a 
90\% chance of being detected if brighter. }
\tablefoottext{a}{Three additional observations (Obsid - Rev = 0148210101 - 0607, 0148210401 - 0701, and 0164561301 - 0960) are available but are totally affected by strong flares: they were thus discarded from the analysis.}
\tablefoottext{b}{Refers only to EPIC MOS data.}
\end{table*}

\subsection{\ch}
The pipeline data (level 2) available in the \ch\ archives were downloaded but not 
re-processed. Further work was done with CIAO 4.2 and CALDB 4.3.1. We first 
defined the source and background regions. The sources regions are circular,
with radii of 5'' (i.e. 10px), while the background regions are chosen as 
surrounding annuli of radii 5'' and 15''. The sources W243 and LBV 1806-20 
have close X-ray companions, so that the annular background regions 
were replaced by nearby circular regions of 5'' radii in these two cases.
We then used the task {\it specextract} to get unbinned spectra of source 
and background regions, as well as their corresponding response matrices. 
When several observations are available, the task {\it combine\_spectra} 
was used to co-add the spectra.

If the source is detected, the flux in the 0.5--8.0\,keV band is then readily 
estimated using an absorbed optically-thin thermal model, with temperature 
fixed to 0.6\,keV whenever less than 100\,cts are recorded and free otherwise, and absorptions being 
always fixed to their known values (see previous section). In case of 
non-detection, the count numbers in the two regions, their areas, and the 
exposure times provide upper limits with 90\% significance using the task 
{\it aprates}\footnote{See http://cxc.harvard.edu/ciao/threads/upperlimit/
 - we used it with alpha=0.99 (the PSF fraction encircled by the source region)
and beta=0.01 (the PSF fraction encircled by the background region).}
which relies on Bayesian statistics to evaluate errors. 
Using the calculated response matrices and a ``fakeit'' run under Xspec, we 
can then transform these count rates into actual fluxes assuming the same 
model as above.

In two cases, the available data were taken using HRC rather than ACIS,
implying that no spectral information is available. For LBV 1806-20, 
two other ACIS datasets are available. We thus only checked that the limit
derived in the same regions using the task {\it aprates} on the HRC data 
was compatible with the ACIS results. For MN53, the HRC data are the sole
available, and the HRC limit on the count rate found with {\it aprates} was 
transformed into actual flux using WebPIMMs and the above model.

Table \ref{ch} lists our results. 
As for \xmm,
we consider to have a secure detection when the source is detected with
at least 4\,$\sigma$ significance.

Finally, a search was made for the cataloged (c)LBVs in the Chandra Source 
Catalog (CSC)\footnote{http://cxc.harvard.edu/csc/index.html}. X-ray sources 
were only found close to GAL 026.47+00.02, W243 and Sher 25. For the 
former, the X-ray source lies at 0.2'', implying detection, but for the last 
two objects, the CSC X-ray sources lie at $>5$'', casting doubt on their 
identification with the LBVs.

\subsection{Proprietary data}
Sher 25 is a special case in our survey since 5 observations exist but 4
of these were still proprietary at the time of the analysis. These additional data
greatly enhance the detection limit, though, as the total exposure time is
then multiplied by 10. The PI of these data, L. Townsley, kindly made
them available to us. The full dataset (the public exposure and the 4
private observations) was reduced and combined using the {{\em ACIS Extract}} software package\footnote{The {\em ACIS Extract} software package and User's Guide are 
available at  http://www.astro.psu.edu/xray/acis/acis\_analysis.html , 
see also \citet{Broos2010}.} 
by P. Broos: we use here an extraction specifically made
at the position of Sher\,25 on the full dataset.

Similarly, a deep analysis of the X-ray emission of MCW 930 was possible
because the PI of the \xmm\ data, L. Bassani, kindly made the observation
available to us before the end of her proprietary period. These data were 
reduced as our other \xmm\ data (see Sect. 3.1). Note that there is also 
a \ch\ dataset (400638 - 7528) for this source, but it has an exposure time of 
only 5ks ; moreover, this cLBV appears on the edge of a CCD, prohibiting 
the derivation of any meaningful flux limit.

It may be worth noting that 2 additional sources (GRS 79.29+0.46 and 
2MASS J16364278--4656207, see Table \ref{cat}) have \ch\ data, but they 
are not yet publicly available. 

\subsection{The Galactic Center survey}
Ten of the cataloged cLBVs lie close to the Galactic Center. A survey of 
this region, using 2Ms of \ch\ time, was reported by \citet{mun09}. This 
paper and its associated on-line resources\footnote{See 
http://www.srl.caltech.edu/gc\_project/xray.html and tables on the journal webpage (CDS tables are incomplete and tables on the Caltech website contain errors).} provide source lists 
and source spectra, as well as sensitivity maps in the 0.5--8.0\,keV 
energy range (see their Fig. 9). We use the sensitivity map where sources brighter 
than the quoted limiting photon flux have $>90$\% chance of being detected 
(the other one available corresponds to a $>50$\% chance). This 90\% 
probability corresponds to $1.65\sigma$ or a logarithmic likelihood of 2.3.

\begin{table*}
\caption{\ch\ imaging observations of (c)LBVs. A `H' after the Obs\_ID indicates HRC data, the off-axis angle (OFA) is the distance to the non-offset aimpoint.}
\label{ch}
\begin{center}
\begin{tabular}{lccccccc}
            \hline\hline
Simbad Name & seqnum & Obs\_ID & OFA & count rate & $F^{\rm obs}_{\rm X}$ & $L^{\rm unabs}_{\rm X}$ & \loglxlb \\
 & & & (\arcmin)& ct\,s$^{-1}$ & erg\,cm$^{-2}$\,s$^{-1}$ & \ergs & \\
\hline
Cl* West. 1 W 243  & 200344 & 5411 & 1.7& 0.00029..61$\pm$0.00015: & $1.7..3.5\times 10^{-15}$: & $1.5..3.1\times 10^{32}$: & $-7.3..-7.0$:\\
                        & 200344 & 6283 & 2.1& & & & \\
GAL 026.47+00.02        & 400603 & 7493 & 4.3& 0.0161$\pm$0.0009 & $(1.73\pm0.10)\times 10^{-13}$ & $(4.3\pm0.2)\times 10^{33}$ & $-5.95\pm0.02$\\  
NAME LBV 1806-20\tablefootmark{a}    & 500599 & 6251H& 0.4& $<$0.0009 & $<3.7\times 10^{-14}$ & $<6.9\times 10^{34}$ & $<-5.0$\\ 
                        & 500042 & 746  & 0.6& $<$0.0011& $<1.0\times 10^{-14}$ & $<2.0\times 10^{34}$ & $<-5.6$\\
NAME Sher 25 star       & 200058 & 633  & 0.6& $<$0.000017& $<1.1\times 10^{-16}$ & $<4.9\times 10^{30}$ & $<-8.8$\\  
                        & 200666 & 12328& 0.6& & & & \\
                        & 200666 & 12329& 0.5& & & & \\
                        & 200666 & 12330& 0.5& & & & \\
                        & 200666 & 13162& 0.5& & & & \\
$[$OMN2000$]$ LS1       & 200132 & 2524 & 5.9& $<$0.00018 & $<1.4\times 10^{-15}$ & $<1.6\times 10^{32}$ & $<-7.1$\\
                        & 200132 & 3711 & 5.9& & & & \\ 
2MASS J18415965 & 400662 & 7552 & 6.2& $<$0.00016 & $<1.3\times 10^{-15}$ & & \\
\hspace*{10mm} --0515409\\
MN53                    & 500140 & 1963H& 7.4& $<$0.002   & $<2.4\times 10^{-14}$ & & \\
MN79                    & 500629 & 6675 & 1.7& $<$0.00025 & $<1.0\times 10^{-15}$ & & \\
MN83\tablefootmark{b}   & 400603 & 7493 & 5.2& $<$0.0002  & $<1.5\times 10^{-15}$ & & \\
\hline
eta Car                 & \multicolumn{4}{c}{see \citet{cor10}} & $0.06..3\times 10^{-10}$ & & $\sim$--5\\
NAME VI Cyg 12          & \multicolumn{4}{c}{see \citet{rau11}} & $2.7\times 10^{-12}$ & $8.2\times 10^{33}$ & $-6.1$\\
\hline	
\multicolumn{4}{l}{\it Galactic Center} & photon fluxes (ph\,cm$^{-2}$\,s$^{-1}$) & & & \\
qF 362                  & & & & $<1.5\times 10^{-6}$ & $<5.6\times 10^{-15}$ & $<7.8\times 10^{33}$ & $<-5.9$\\
GCIRS 34W               & & & & $(9.8\pm0.2)\times 10^{-6}$ & $7.5..8.1\times 10^{-14}$ & $1.0..1.1\times 10^{33}$ & $\sim-6.1$\\
Pistol Star             & & & & $<1.5\times 10^{-6}$ & $<5.6\times 10^{-15}$ & $<7.8\times 10^{33}$ & $<-5.9$\\
WR102ka                 & & & & $<2\times 10^{-6}$ & $<7.4\times 10^{-15}$ & $<5.0\times 10^{33}$ & $<-6.4$\\ 
GCIRS 16NE              & & & & $<3.7\times 10^{-6}$ & $<1.4\times 10^{-14}$ & & \\
GCIRS 16C               & & & & $<3.7\times 10^{-6}$ & $<1.4\times 10^{-14}$ & $<9.1\times 10^{33}$ & $<-5.5$\\
GCIRS 16SW              & & & & $<3.7\times 10^{-6}$ & $<1.4\times 10^{-14}$ & & \\
GCIRS 16NW              & & & & $<3.7\times 10^{-6}$ & $<1.4\times 10^{-14}$ & $<9.1\times 10^{33}$ & $<-5.5$\\
GCIRS 33SE              & & & & $(1.41\pm0.02)\times 10^{-5}$& $(1.42\pm0.02)\times 10^{-13}$ & $(1.23\pm0.02)\times 10^{33}$ & $-6.244\pm0.006$\\ 
2MASS J17460562 & & & & $<1.2\times 10^{-6}$ & $<4.4\times 10^{-15}$ & $<6.2\times 10^{33}$ & $<-6.0..6.4$\\
\hspace*{10mm} --2851319\\
\hline	
\end{tabular}
\end{center}
\tablefoot{Column 1 provides the (c)LBV name, 
columns 2 to 4 the details of the observation used, and the next columns
present the derived count rates, observed fluxes, and absorption-corrected luminosities in the 
0.5--8.0\,keV energy range. For the detected objects, the quoted errors 
correspond to 1-$\sigma$; for non-detections, the quoted limits correspond 
to a 90\% chance of being detected were the source brighter. }
\tablefoottext{a}{One additional observation (seqnum=500598, obsid=6224) covered the region containing this cLBV, but at a very large off-axis angle: the presence of a neighbouring bright pulsar prevents any detection at this position.}\\
\tablefoottext{b}{Two additional observations (seqnum=500759, obsid=7630 and 9754) covered the region containing this cLBV, but the source lies on the CCDs outer edges. }
\end{table*}

\section{Results}
In this section, we present the results of this survey. We begin by 
presenting the detections, and then continue with the doubtful detections and the non-detections.
Whenever the X-ray emission of the object has been discussed in the past,
it is clearly mentioned.

\subsection{Detections}

\subsubsection{eta\,Car}
Of all LBVs, eta\,Car is certainly the most spectacular, luminous, and 
best studied object... but not necessarily the best understood one \citep{dav97}!
Its fame comes from the two impressive brightenings observed in the 19th
century (the ``Great Eruption'' in mid-19th century and the smaller event
of 1890). Since then, there has been a lot of discussion about the 
variability of this star, and its potential recurrence timescale.
A period of 5.5\,yr was finally identified by \citet{dam96}, and
it is now detected from radio to X-ray energies.

Though there is no direct evidence for a companion, this timescale is
often interpreted as the orbital period of an eccentric binary 
\citep[for a review see ][]{cor11}. 
Indeed, a colliding-wind binary (CWB) can easily explain many properties
of the observed variability (e.g. the nebula's changes are related to
photo-excitation variations linked to the collision effects), though
an exact modelling reproducing all observed details is generally 
difficult \citep[see e.g. in X-rays the work of][]{par11}. 

In the X-ray domain, it is quite remarkable that eta\,Car has been one of the first massive stars 
detected at these energies \citep{sew79}, soon after a few Cygnus OB2 stars 
(see Sect. 4.1.3 below). It has been intensively studied since then, and the observed
hardness of the emission and its variability properties have indicated 
a strong similarity with the well-known CWB WR\,140 \citep[for recent campaigns on these objects, see ][]{cor11,wil11}. An extensive {\it RXTE} monitoring 
\citep{cor05,cor10} notably showed that the 
X-ray emission appears rather stable around apastron, but rises before
periastron, as the two stars come closer together. At periastron,
an abrupt decrease occurs, and this short minimum state is often interpreted
as an occultation or eclipse (combined with a switch-off, \citealt{par11}) 
of the CWB emission. The X-ray lightcurve is however not strictly periodic. 
The 2003.5 event was more luminous than the 1998 event, and the recent
2009 event displayed a lower and softer flux before periastron, as well
as a shorter minimum state \citep{cor05,cor10}. This may be linked to
changes in the wind density, through variations of the wind velocity, 
mass-loss rate or both \citep{cor10}. The observed luminosities (in the 
2--10.\,keV range) vary from $2\times 10^{35}$\,\ergs\ at maximum to 
values 50 to 100 times smaller at minimum \citep{ish99,cor10}. 
As the actual value of the absorbing column and bolometric luminosity 
are still debated, it is difficult to give accurate \loglxlb, but 
eta\,Car may certainly reach values of --5. An example of its X-ray spectrum
is shown in Fig.\,\ref{specgood}

\subsubsection{W 243  }
Using two \ch\ observations, \citet{cla08} studied the X-ray emission 
of stars in the Westerlund 1 cluster. These authors mention a ``weak 
detection'' for W243: their Table 4 lists $7\pm6$ net counts for this
object, corresponding to a luminosity of about 10$^{32}$\,\ergs\ according to their 
count-to-flux conversion rate\footnote{see their Table 4; this probably 
corresponds to a conversion towards absorption-corrected flux, though 
it is not clearly indicated in the paper.}.

We re-analyzed the same \ch\ data, with a different technique (see 
Sect. 3.2) and we also take a look at the many \xmm\ datasets encompassing
the source. 

\begin{figure}
\begin{center}
\includegraphics[width=5.3cm,height=7cm,angle=-90, bb=67 43 570 772,clip]{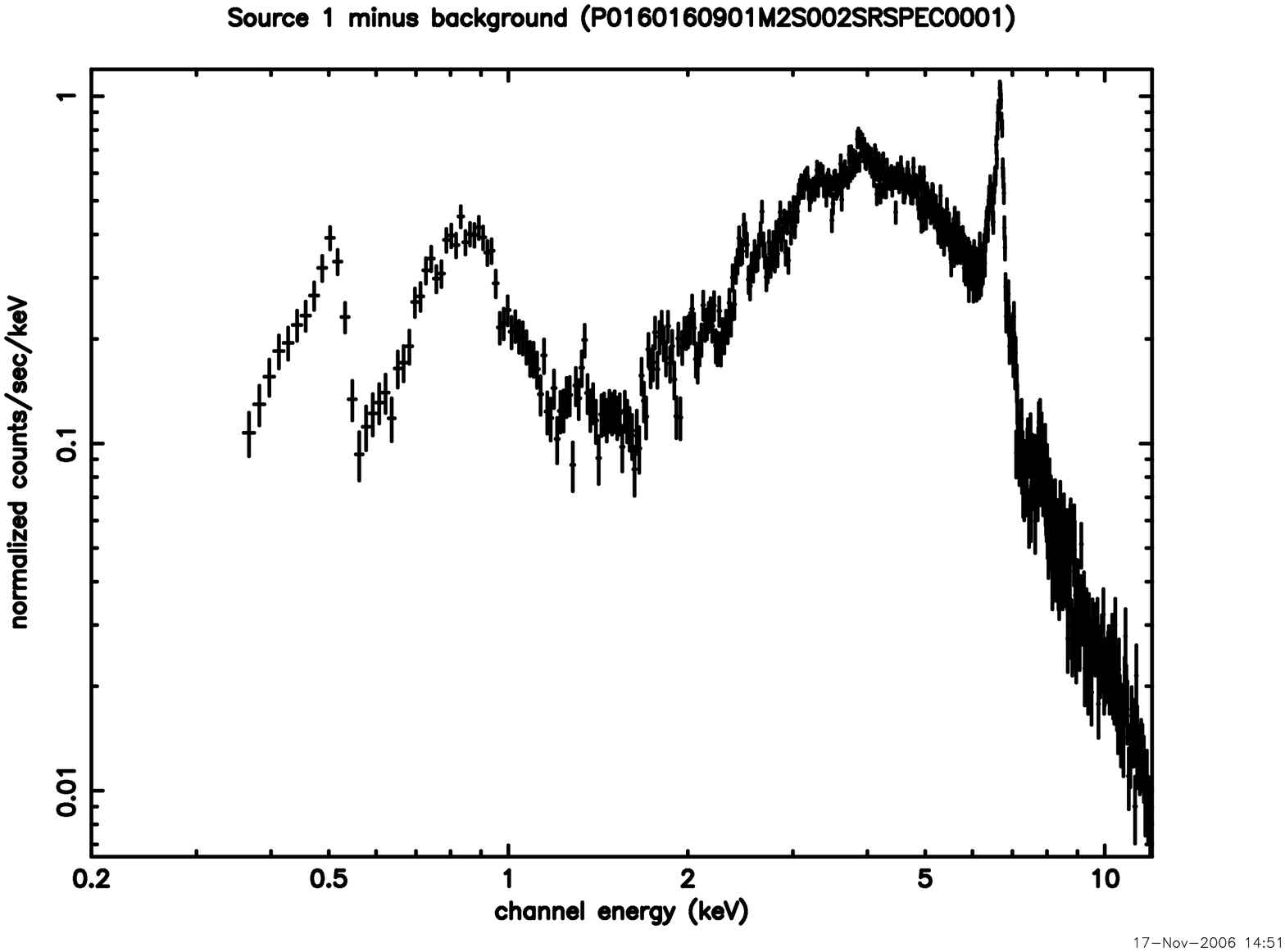}
\includegraphics[width=7cm,height=5.3cm]{w243.ps}
\includegraphics[width=5.3cm,height=7cm,angle=-90]{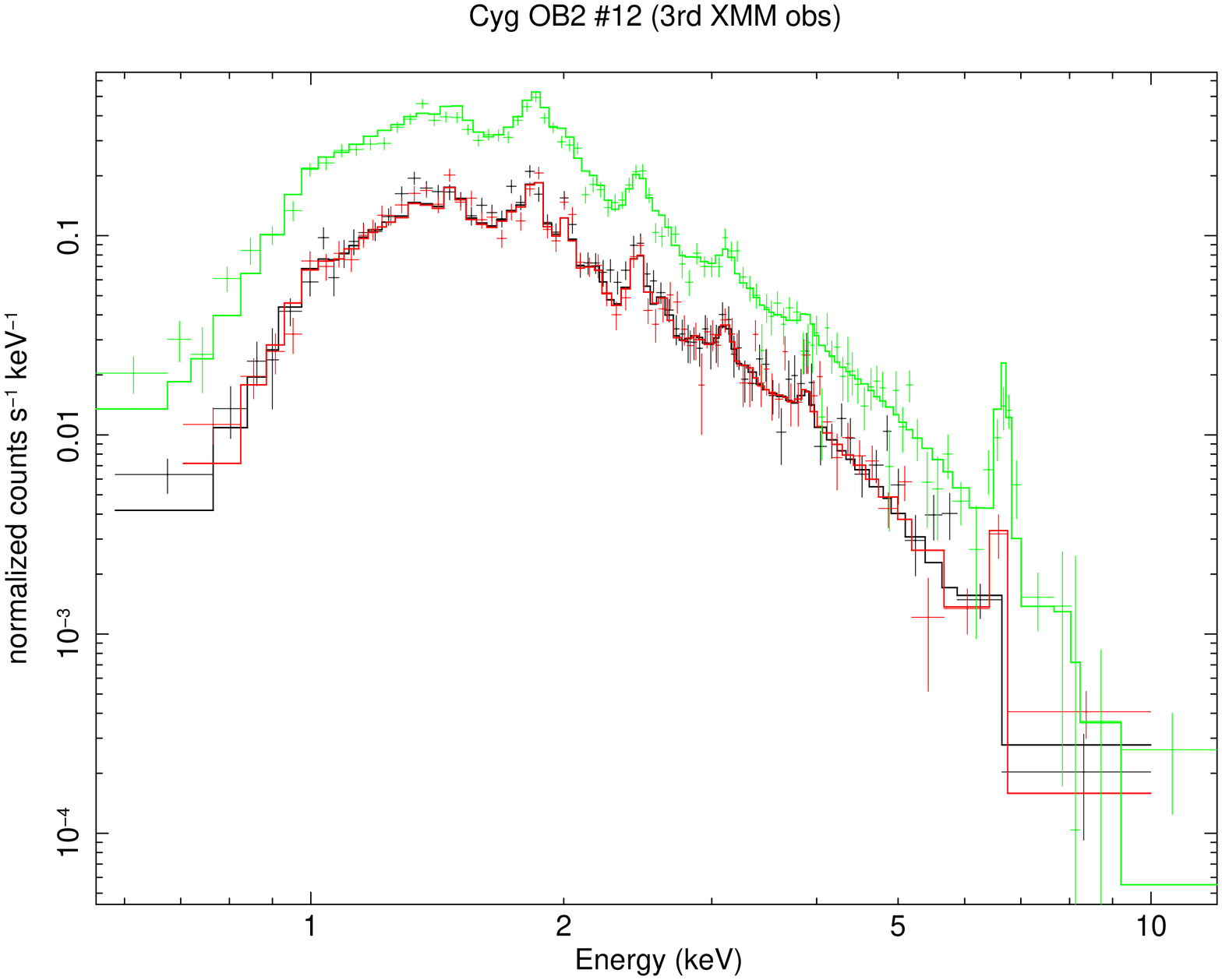}
\includegraphics[width=7cm,height=5.3cm]{spec_gal26.ps}
\caption{X-ray spectra of the detected LBVs (from top to bottom: eta Car - from the 2XMM database, W243, Cyg OB2 \#12 - from \citealt{rau11}, and GAL 026.47+00.02), along with their best-fit model (see text for detail). The eta Car spectrum corresponds to XMM-MOS2 data taken at the June 2003 maximum, which are dominated by the soft emission of the outer ejecta below 1.5\,keV and by colliding-wind emisson above that limit \citep{ham07}. The W243 spectrum corresponds to the 5'' extraction region, and was re-binned by a factor of 10. Note the strong iron line at 6.7\,keV for eta Car and Cyg OB2 \#12.}
\label{specgood}
\end{center}
\end{figure}

As the \ch\ data show that there is a source at 7.5'' south
of W243 (CXOU J164707.6--455235), we do not use an annulus to estimate the background 
but rather a nearby circle devoid of sources (see Sect. 3.2). 
When extracting the data at the position of W243 in both \ch\ datasets and 
combining them, we found 34 net counts for the source if the
source region has 5'' radius, and 15 net counts if we reduce the 
extraction radius to 3.4'' (to avoid contamination by the neighbour). 
This leads to count rates of $0.00061\pm0.00015$ and 
$0.00029\pm0.00010$\,cts\,s$^{-1}$, respectively. 
Both results are compatible with each other (e.g. count rates are
within $3\sigma$), but only the first case constitutes a formal detection with 
our definition (see Sect. 3.1 and Fig. \ref{specgood}). The region used in the first case is also the one most prone 
to contamination from the neighbouring object. Indeed, changing 
the extraction radius to 3.4'' yields similar results for other \ch\
targets.
We therefore consider the count rate as uncertain, advocating 
for a ``weak detection'' status at best.

All \xmm\ data were taken with the medium filter except for the first 
two datasets (0404340101 - 1240 and 0311792001 - 1243) which we therefore
exclude from the mosaicking process.  There is no trace of the 
source due south of W243. In fact, this source and W243 both lie on the 
PSF wing of the pulsar PSR J1647--4552 which is situated at 31'' east 
of W243. The presence of this bright X-ray source severely limits the sensitivity 
achieved at the position of W243. Therefore, since the LBV seems 
undetected in the \xmm\ data, an upper limit on its flux was derived 
by adding a fake source, with a similar PSF, at the position of the 
LBV\footnote{The astrometric errors of the \xmm\ datasets 
are negligible as the error on the pulsar's position, found by 
comparing the Simbad coordinates with those in the detection list 
of the combined mosaic amounts to only 0.2''.}. We found that an X-ray source $\sim400$ 
times fainter than the pulsar's emission would be barely discernible, 
implying an upper limit on the EPIC count rate of 0.0017\,cts\,s$^{-1}$ 
(Table \ref{xmm}). Note that, as the pulsar was brighter in the two
excluded observations, we obtained a better sensitivity limit here
than when considering the whole \xmm\ set. The limit derived from \xmm\
data is quite similar to the values found using the weak
\ch\ detection, which means that \xmm\ is on the verge of detecting W243. 
However, the \ch\ observatory appears better suited for the purpose of 
studying W243 - deeper observations would help resolve the 
uncertainties on its properties.

In any case, the modest X-ray emission level of 10$^{32}$\,\ergs\ observed 
in W243 is rather typical of OB stars (\citealt{har79,ku79}, see also recent cluster studies of \citealt{san06,ant08,naz11b}), as well as 
its \loglxlb\ of --7 \citep{pal81,ber97,naz09,naz11a}. This leads to two possibilities.
On the one hand, W243 could be a single LBV with a wind comparable to that 
of a ``normal'' O-type  star. While its mass-loss rate is indeed similar 
to that of O-type stars, the wind terminal velocity of W243 was found 
to be only 165\,\kms, i.e. one order of magnitude lower than in O-stars 
\citep{ri09}. Such a slow wind casts doubt on the ``LBV wind-shock model'' 
scenario, as the (potential) intrinsic shocks will be rather weak to 
generate enough X-rays, compared to O-stars. On the other hand,
W243 could actually be a multiple system composed of an LBV and an O-type star 
whose intrinsic high-energy emission is the actual source of the recorded
X-rays, rather than the LBV itself. 

\subsubsection{NAME VI Cyg 12 = Cyg OB2 \#12}
With $M_V \simeq -10$, Cyg\,OB2 \#12 is considered to be one of the 
optically brightest stars in our Galaxy (e.g.\ \citealt{massey}). 
 Despite this high luminosity placing it above the Humphreys-Davidson 
limit, Cyg OB2 \#12 lacks some of the LBV characteristics, such as 
temperature changes, enrichment, or traces of circumstellar material
\citep{cla11}. This 
object was one of the very first massive stars detected as an X-ray 
source during an {\it EINSTEIN} observation of Cyg\,X-3 \citep{har79}. 
This star has subsequently been observed by all major X-ray satellites. 
Most recently, \citet{rau11} performed an in-depth monitoring of this object using
six \xmm\ pointings obtained in October-November 2004 and April-May 
2007. The EPIC spectra of this star (Fig . \ref{specgood}) appeared quite hard: it was fitted 
with an absorbed two-temperature thermal plasma model consisting of 
plasma at temperatures of $kT \simeq 0.75$ and 2\,keV (where the strength of the second component is quite similar to that of the first one, not much smaller). The average 
absorption-corrected X-ray flux in the 0.5--8.0\,keV domain was 
$2.4 \times 10^{-11}$\,erg\,cm$^{-2}$\,s$^{-1}$, which corresponds 
to an X-ray luminosity of $8.2\times 10^{33}$\,erg\,s$^{-1}$ 
assuming a distance of 1.7\,kpc and $\log[L_{\rm X}/L_{\rm BOL}]=-6.1$ (using reddening,
distance and bolometric luminosity of Table \ref{cat} rather than 
the values used by \citealt{rau11} - though the final \loglxlb\ value
are then equal). 

These temperatures, especially the second one, cannot be 
explained by shocks in a slow (150\,km\,s$^{-1}$) moving wind as 
proposed by \citet{KC}  and more recently by \citet[400\kms]{cla11}. However, \citet{SL} reported a variable 
absorption bluewards of H$\alpha$ that they tentatively interpreted 
as a blueshifted H$\alpha$ absorption arising in an expanding shell 
at a velocity of 1400\,km\,s$^{-1}$. \citet{LHSW} actually inferred 
this velocity to correspond to the wind velocity of Cyg\,OB2 \#12. 
An expanding shell of even higher velocity (around 3100\,km\,s$^{-1}$) 
was reported by \citet{WZ}. A faster wind than inferred by \citet{KC}
is therefore possible, though the tremendous interstellar reddening 
unfortunately hampers a reliable determination of the star's wind 
velocity. 

On the other hand, the EPIC data further revealed X-ray flux variability (with 
an amplitude of $\sim 10$\%) on timescales ranging from a few days 
to a few weeks, as well as larger variations (37\%) on timescales 
of years \citep{rau11}.  The combination of these properties (strong, 
variable X-ray emission, including a strong high-temperature component) 
suggest that Cyg\,OB2 \#12 could be either an eccentric colliding 
wind binary (with a secondary star featuring a fast wind) or a single 
star featuring a strong magnetic field that confines the stellar wind 
near the magnetic equator.

\subsubsection{GAL 026.47+00.02        }
GAL 026.47+00.02 was detected by \ch\ as a rather faint X-ray source (only
$\sim 300$ counts recorded in 20\,ks). This X-ray source is well isolated, 
thereby giving credit to its association with the cLBV. 

A rough (in view of the S/N) X-ray spectrum could be extracted for
this object (Fig. \ref{specgood}). A fit by an absorbed optically thin thermal plasma model
yields a temperature of $0.47\pm0.11$\,keV and a normalization factor
of $(5.7\pm0.4)\times 10^{-4}$\,cm$^{-5}$ when the absorption is fixed 
to the ISM value. The resulting $\log[L_{\rm X}/L_{\rm BOL}]=-6$ is similar to that of
Cyg\,OB2 \#12, though the former object appears somewhat softer.
Such a hard and bright X-ray emission suggests that there are colliding
winds or magnetically confined winds in the system. However, follow-up 
observations (e.g. to test whether the X-ray source varies, and how) would 
be required to further constrain the nature of GAL 026.47+00.02.

\subsection{Doubtful detections}

\subsubsection{GCIRS 34W               }
This LBV appears at 0.3'' of the X-ray source CXOUGC J174539.7--290026 in the 
Galactic Center survey \citep{mun09}. However, the X-ray source does not 
appear point-like but rather looks like a diffuse emission region. This 
casts doubt on the identification of the source of the recorded X-rays with 
the sole LBV.

\begin{figure}
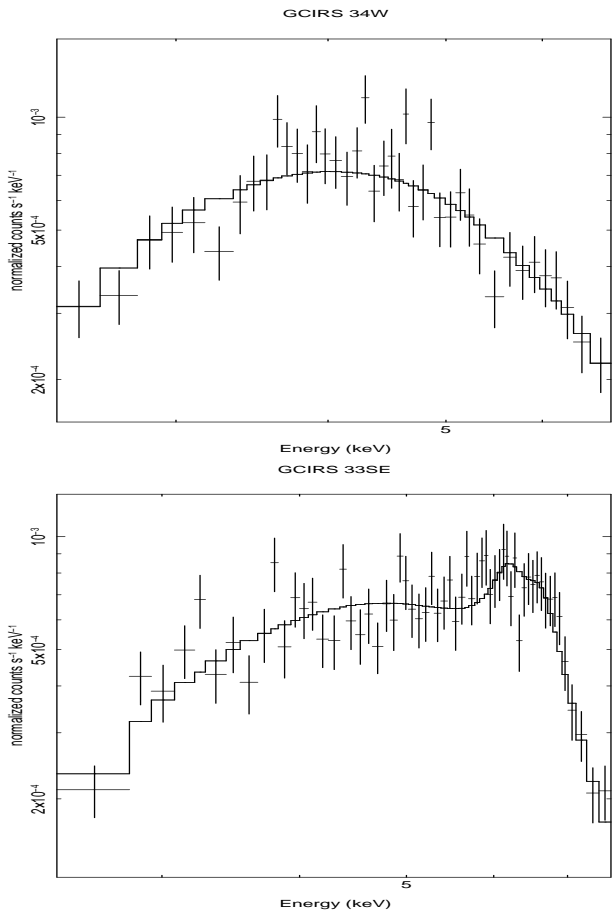

\begin{center}
\includegraphics[width=8cm,height=6cm]{spec_34W.ps}
\includegraphics[width=8cm,height=6cm]{spec_33SE.ps}
\caption{X-ray spectra of the X-ray sources apparently associated with GCIRS 34W and GCIRS 33SE, along with their best-fit model (see text for detail). Note the absence of lines, especially in comparison with Fig. \ref{specgood}.}
\label{specbad}
\end{center}
\end{figure}

This conclusion is supported by the appearance of the X-ray spectrum (Fig. \ref{specbad}). An absorbed 
optically-thin thermal plasma provides a rather good fit but only with extremely high temperatures: $kT>53$\,keV,
$norm=(1.09\pm0.05)\times 10^{-4}$\,cm$^{-5}$, and an absorption fixed
to the ISM value yield $\chi^2=1.5$. An absorbed powerlaw is clearly
better: $\Gamma=0.95\pm0.16$, $norm=(1.1\pm0.2)\times 10^{-5}$\,cm$^{-5}$, 
and an absorption fixed to the ISM value yield $\chi^2=1.1$. Both models
yield quite similar flux values, though (both are given in Table \ref{ch}).

\subsubsection{GCIRS 33SE              }
The cLBV is at 0.3'' of the X-ray source CXOUGC J174540.0--290030 of the 
Galactic Center survey \citep{mun09}. This source was found to vary
in flux by two orders of magnitude (from photon fluxes of $1.6\times 
10^{-6}$\,ph\,cm$^{-2}$\,s$^{-1}$ to $2.3\times 
10^{-4}$\,ph\,cm$^{-2}$\,s$^{-1}$). 

The mean spectrum appears extremely hard , with all photons at energies 
larger than 2\,keV (Fig. \ref{specbad}). It could not be fitted by an absorbed powerlaw,
an absorbed optically-thin thermal plasma, an absorbed bremsstrahlung,
or an absorbed disk blackbody model. In fact, a good fit was only achieved
by adding a flat powerlaw to a line (either a gaussian or a relativistic one):
for the latter case (model {\it diskline} in Xspec), we found
$\Gamma=-0.08\pm0.12$, $norm_{\rm PL}=(2.2\pm0.6)\times 10^{-6}$\,cm$^{-5}$,
$E_{\rm line}=6.41\pm0.05$\,keV, $i>77^{\circ}$, and
$norm_{\rm line}=(4.8\pm0.4)\times 10^{-6}$\,cm$^{-5}$.

This type of spectrum and variations are rather atypical of massive stars
(even considering the ``hard'' colliding-wind systems) and would better fit 
in the context of X-ray binaries. Therefore, we doubt that the survey has 
detected the true emission of GCIRS 33SE.

\subsection{Non-detections}
\subsubsection{P Cyg                   }
The \ros\ data of P\,Cyg were analyzed by \citet{ber00} and \citet{osk05a}. 
However, these X-ray data suffer from strong contamination by the intense UV emission from 
this hot star, which implies most probably a non-detection of P\,Cyg in
this dataset. Limiting (absorption-corrected) luminosities of $9\times 10^{30}$\,\ergs\ and 
$2\times 10^{31}$\,\ergs\ in the \ros\ range (0.1-2.5\,keV) were reported by \citet{ber00} 
and \citet{osk05a}, respectively. This corresponds to \loglxlb\ below 
$\sim -8$ (considering the same soft energy range).

The new \xmm\ observation was taken with the thick filter, to avoid any
UV contamination. In the new data, three faint X-ray sources are clearly detected towards the east of 
P\,Cyg. They are at distances of 28'' (for two of them) and 62'' (for
the last one of the trio). This is well above the potential astrometric 
errors and the PSF size, i.e. these source are too distant for being 
reasonably associated with the LBV.

Definitely, there is no X-ray source at the position of P\,Cyg. The upper 
limit on the flux is more than an order of magnitude lower than the \ros\ 
limit though it corresponds to a harder energy range (0.5--8.\,keV for \xmm\
vs 0.1--2.5\,keV for \ros). The signal recorded by \ros\ was thus clearly 
dominated by UV contamination.

\subsubsection{AG Car                  }
The field of AG\,Car looks very similar to that of P\,Cyg: only faint X-ray 
sources scattered everywhere, without anything at the position of the LBV. 
Here, the closest X-ray source lies at nearly 1' of AG\,Car, and there is 
thus no doubt on the non-detection of the LBV. 

Note that Hen 3-519 is rather close to AG\,Car, but no constraint on its 
X-ray flux can be derived from these data, as it is out of the field-of-view 
of the MOS cameras and just on the edge of a pn CCD.

\subsubsection{HD 160529               }
The \xmm\ observation appears more densely populated than in the cases 
of P\,Cyg and AG\,Car, but the closest X-ray source appears at 56'' of 
HD\,160529 and, again, the LBV clearly remains undetected.

\subsubsection{qF 362                  }
The deep Galactic Center Survey data do not show any X-ray source in the 
vicinity of qF 362. The closest X-ray source, CXOUGC J174616.6--284909,
lies at more than 18'', which is clearly too far away from the LBV 
considering the narrow PSF of \ch.

\subsubsection{GAL 024.73+00.69        }
A strong flare affected the \xmm\ observation, reducing the useful exposure 
to only 3.5\,ks. This indeed limits the sensitivity of the observations:
only 7 X-ray sources were detected in the field-of-view and none appears
close of the LBV (they are at separations $>7$'). 

\subsubsection{Pistol Star             }
Before the publication of the full Galactic Center survey, 
\citet{mun06} and \citet{osk05b} analyzed the then available \ch\ exposures 
of the region, and they report a limiting \loglxlb\ of $-8.2$ and $-8$, 
respectively. 

The deep Galactic Center Survey data do not show any X-ray source in the 
vicinity of Pistol star. The closest X-ray source, CXOUGC J174615.2--284957,
lies at about 6''. The derived limit on the \loglxlb\ is --6, which may seem
at odds with previous reported results, though the survey of \citet{mun09}
uses more datasets. Though no detail on the exact derivation of the
\loglxlb\ limit is provided in these previous papers, it is interesting to 
note that such limits are found if one uses the {\it observed} flux, not the 
{\it absorption-corrected} one: as the absorption in the direction of the
Galactic Center is high, this explains the two order of magnitude apparent
discrepancy in \loglxlb. However, if one wants to compare objects scattered 
all over the Galaxy, it is important to correct the luminosities for
the interstellar absorption - indeed, the `canonical' $\log[L_{\rm X}/L_{\rm BOL}]=-7$ of O-stars
only applies if this correction is done. Unfortunately, this results in
a much weaker constraint on the X-ray emission of the Pistol star.


\subsubsection{HD 316285               }
Data in the vicinity of this cLBV were collected by \xmm\ during 3
observations. For 0112970201 - 0145, it appears so far off-axis that it is
out of the field-of-view of MOS2. In addition, there are a lot of flares
in 0205240101 - 0956. This probably explains why the limiting count rate
derived with only the 0112970101 - 0145 dataset is more constraining than
the limit derived when considering all 3 datasets. We therefore quote 
in Table \ref{xmm} the former value only. 

\subsubsection{GRS 25.5+0.2            }
This cLBV was observed quite far off-axis (10') in a relatively short (7.5\,ks)
exposure: the sensitivity is thus quite shallow, and the derived limit on 
the count rate is accordingly high. The closest X-ray source appears at 4'
of the cLBV, excluding any confusion problems.

\subsubsection{WR102ka                 }
The deep Galactic Center Survey data do not show any X-ray source in the 
vicinity of WR102ka. The closest X-ray sources, CXOUGC J174616.6--290115
and J174618.3--290204, lie at about 28''. 

\subsubsection{LBV 1806-20        }
This cLBV is situated 14'' east of the bright pulsar SGR 1806-20.
While this separation is not a problem for \ch, the cLBV is situated 
on the PSF wing of this bright pulsar in \xmm\ instruments. We did 
a mosaic with all available \xmm\ observations, except the last one 
(0604090201 - 1785) which was taken with a different filter, but the 
produced sensitivity map has no actual meaning. The situation is even worse
than for W243 as the separation is smaller: the detection 
is mostly hampered by the closeness of the pulsar rather than the 
available effective area. Therefore, since the cLBV seems undetected, 
we estimate the true detection limit by adding a fake source, with a 
similar PSF, at the position of the cLBV\footnote{Note that the astrometric 
errors of the \xmm\ datasets are very limited as the error on the 
pulsar's position, found by comparing the Simbad coordinates with those 
in the detection list of the combined mosaic amounts to only 0.2''.}. 
We changed its amplitude until it was not detectable. We found that an 
X-ray source 100 times fainter than the pulsar's emission would be barely 
discernible: this constitutes our flux limit for \xmm, and corresponds 
to an EPIC count rate of 0.01\,cts\,s$^{-1}$ (Table \ref{xmm}). 

For \ch, there is no confusion problem, but the pulsar left a readout trail
in the ACIS data. The background was thus not measured in an annulus, but rather taken in
a circle of the same size as the source region and lying on the same readout 
trail. This resulted in the limit quoted in Table \ref{ch}. The same regions 
were indeed used for the HRC data, for coherence reasons. Note that the cLBV 
is formally detected at $1\sigma$ significance (i.e. not an actual detection 
with our criteria, see Sect. 3.1) in the ACIS data, but this 
is clearly an artifact due to the readout trail. The \ch\ observations 
are better suited to disentangle the pulsar from the cLBV, but they
are unfortunately not very sensitive, thereby not improving the \xmm\ results.

\subsubsection{Sher 25       }
Using a single observation of NGC\,3603, \citet{mof02} reports a non-detection
of Sher\,25, without any detail on the limiting X-ray flux.

The additional recent observations enable us to enhance the sensitivity 
in the region by an order of magnitude, resulting in a tight upper limit
on the X-ray emission of Sher25. Note that the source is formally detected
with $1\sigma$ significance (i.e. not an actual detection with our criteria, 
see Sect. 3.1) as there are $3.3\pm3$ net counts in the source region.

\subsubsection{$[$OMN2000$]$ LS1       }
Situated in the outskirts of the massive star-forming region W51A,
LS1 shows a formal 1$\sigma$ ``detection'' (i.e. not an actual detection  
with our criteria, see Sect. 3.1), with an upper limit on the observed 
flux of $1.4\times 10^{-15}$\,erg\,cm$^{-2}$\,s$^{-1}$.

\subsubsection{GCIRS 16NE, 16C, 16SW, 16NW              }
The deep Galactic Center Survey data do not show any X-ray source at the 
position of these objects. The closest X-ray sources are GCIRS 33SE and 
34SW (see above), as well as CXOUGC J174540.1--290025, J174540.0--290028 
and J174539.7--290029. For all sources, we use the same absorbing column
to derive the limiting flux.

This region of the survey is the one observed with the largest sensitivity,
which explains why the limiting photon fluxes are so low. However, this
is also the most crowded region of the survey: the derived limits may
thus be somewhat optimistic, although it is difficult to quantify by how much.

\subsubsection{2MASS J17460562--2851319 }
The deep Galactic Center Survey data do not show any X-ray source in the 
vicinity of this target. The closest X-ray source, CXOUGC J174604.4--285139,
lies at about 17''. 

\subsubsection{MCW 930}
The position of this cLBV falls in a gap of the pn camera. Only MOS data were
therefore used to derive an upper limit on its X-ray emission. The closest
X-ray source appears more than 3' away from the cLBV. There is therefore no
doubt on the non-detection of this object.

\subsubsection{2MASS J17110094--3945174 }
Situated in a field containing only a few faint X-ray sources, this object
remains undetected in the \xmm\ observation. While a flare affected the 
exposure, especially the pn data, the target was observed nearly on-axis,
leading to a rather deep upper limit. Unfortunately, as for the other
objects of \citet{wac10} and \citet{gva10}, no information on the other
properties of the object are available, preventing the derivation of a 
\loglxlb.

\subsubsection{2MASS J18415965--0515409 }
No bright X-ray source is present in the field-of-view of the \ch\ 
observation, and no X-ray source lies closer than 1' away from 
2MASS J18415965--0515409. A flux limit of 
$1.3\times 10^{-15}$\,erg\,cm$^{-2}$\,s$^{-1}$ was derived.

\subsubsection{2MASS J16461734--4508478 }
This source is undetected in the \xmm\ observation. The sole X-ray source of
the field-of-view appears at 3.6'. As the target was observed not far from
the on-axis position (off-axis angle=2.6'), a good constraint could be put 
on its count rate.

\subsubsection{2MASS J18022233--2238002 }
The region encompassing 2MASS J18022233--2238002 was observed three times
by \xmm. However, two of these datasets (0135743001 - 0600 and  0135743101 
- 0607) have a bug in their attitude files, which could not be corrected 
(even with the help of the \xmm\ helpdesk). Therefore, no mosaicking
was possible for this target. We instead used a reduction process similar 
to the one applied for \ch\ data: we first derived for each dataset the 
number of counts appearing in the source and background regions by extracting 
data in adequate zones (2 neighbouring circles of 30'' radii), then we 
found the actual exposure times for each dataset at the position of the target
using individual exposure maps, and finally we use the CIAO task {\it aprates}
to get the 90\% detection limit on the count rate. However, the two 
additional datasets did not improve the upper limit found when using only 
the 0135742801 - 0600 exposure, probably because of the flares affecting 
the other Rev. 0600 data and the fact that the target appears in a gap 
of the pn camera for Rev. 0607. Table \ref{xmm} therefore quotes the value 
derived using only the first dataset.

\subsubsection{MN53  }
This object was observed by both \xmm\ and \ch. Both observations were
centered on the bright 1RXS J170849.0--4009, and the field contains a few
additional faint sources (detected only with \xmm). MN53 remains undetected, 
 the closest X-ray source appearing at 70'' in the \xmm\ data. 
Note that the pn data are limited to a small area around the main target, 
so that the flux limit was derived using only MOS data.
In the \ch\ data, MN53 is formally detected at $1\sigma$ significance 
(i.e. not an actual detection  with our criteria, see Sect. 3.1) and the 
flux limit derived on this HRC observation is about 20 times
larger than that found from the \xmm\ data of the same region.

\subsubsection{MN79  }
MN79 appears in a sparsely populated region, containing only faint X-ray 
sources. The closest one in the \xmm\ data lies at 47'' of the target and
at 15'' in the \ch\ data, but the target itself remains undetected.
The two observatories provided similar flux limits.

\subsubsection{MN83  }
Situated on the other side of the aimpoint compared to GAL 026.47+00.02,
MN83 shows a contrasting behaviour by remaining totally undetected. 
The closest X-ray source is at $\sim30$'', and the limiting count rate
derived for MN83 is about two orders of magnitude below that of
GAL 026.47+00.02. 

\section{Discussion}
This first LBV survey in the X-ray domain covers half of the known LBVs 
or candidates (or two-thirds of the catalog of \citealt{cla05}). Despite 
the heterogeneity of the dataset, the derived limits in observed fluxes are 
remarkably similar, lying between $10^{-15}$ and $10^{-14}$\,erg\,cm$^{-2}$\,s$^{-1}$.
However, as the distance and reddening of the surveyed objects vary a lot,
so do the constraints on the X-ray luminosity of LBVs: the derived limits 
span more than four orders of magnitude, from $\sim10^{30}$ to $\sim7\times 
10^{34}$\,\ergs.

Of all surveyed objects, only four are detected in X-rays, 
corresponding to a detection fraction of one out of four in the sample of 
confirmed LBVs, one out of 9 for the Clark's catalog, or one out of 17 
for the full sample of (c)LBVs. Besides the two well-known cases 
of eta\,Car and Cyg\,OB2\,\#12, GAL 026.47+00.02 and W243 were found 
to emit some X-rays. Two additional cases are formally detected, but the 
properties of their X-ray emission cast doubt on the identification
of the high-energy source with the (c)LBVs.

With the results of our survey at hand, what can be concluded about
the X-ray emission of LBVs?

First, LBVs, as a class, are clearly not bright X-ray emitters. Moreover,
it seems unlikely that LBVs possess an intrinsic X-ray emission
of a level comparable to that of O-type stars, which are moderate X-ray 
emitters. While O-stars display a tight relation between their bolometric 
and X-ray luminosities ($\log[L_{\rm X}/L_{\rm BOL}]\sim-7$ with dispersions down to 0.2\,dex for 
clusters, see clusters studies of \citealt{san06,naz11b} and discussions in \citealt{naz09,naz11a}), five 
(c)LBVs in our sample have $\log[L_{\rm X}/L_{\rm BOL}]<-8.2$, P\,Cyg even showing no emission at 
our sensitivity limit of $\log[L_{\rm X}/L_{\rm BOL}]-9.4$! This indicates a significantly 
weaker X-ray emission, compared to O-stars, for the LBVs as a class.
A rather similar conclusion was found for Wolf-Rayet stars, especially 
those of the WC type \citep{osk03}. This could have two causes: either 
the local absorption
by the wind is too high to let any X-rays generated in the deep inner layers 
escape, like for WC stars, or the wind velocity is too small (a few hundred 
\kms, as is often observed in LBVs) to produce hot plasma through
intrinsic wind shocks. While both may well explain the observed lack 
of X-rays in LBVs, it is unfortunately difficult to accurately constrain the 
wind velocity (see e.g. the case of Cyg\,OB2\,\#12 discussed above)
or absorption  (see e.g. the case of eta\,Car, \citealt{cor11}) whatever the phase of LBV activity.
Both hypotheses (and a combination of the two) therefore remain viable 
at the present time.

If LBVs are intrinsically X-ray faint, the few cases of X-ray detections 
should have an extrinsic cause. The most obvious candidate is binarity.
It is indeed well known that wind-wind collisions in massive binaries 
(i.e. O+OB or WR+OB) can produce bright, hard and variable X-ray emission 
\citep[for a review, see][]{gud09}. This is certainly 
the case of the Galactic LBV eta\,Car (see above) as well as of the SMC 
LBV HD\,5980 \citep[and references therein]{naz07}: the phase-locked 
variations of their bright and hard X-ray emission are a definite 
proof of colliding winds. Such a process has also been strongly advocated 
for Cyg\,OB2\,\#12 in view of its spectral properties and its luminosity 
changes (see Sect. 4.1.3 above), though a direct evidence for periodicity in the 
variations is still lacking. In addition, the bright and hard X-rays 
of GAL 026.47+00.02 are again reminiscent of a wind-wind collision but 
additional X-ray data are needed to unveil the variability properties 
of the X-ray emission and ascertain the nature of this cLBV (Sect. 4.1.4).
Finally, W243  does not show any sign of colliding wind X-rays, but its 
X-ray emission may well be that of an O-type companion (see Sect. 4.1.2 above).
For the four detected (c)LBVs, binarity constitutes a plausible explanation,
even if other processes, such as magnetically-confined winds, cannot
be formally ruled out.

Does that mean that other (c)LBVs are single, as they are not X-ray bright?
It must be recalled that both \ch\ and \xmm\ have shown that
X-ray bright CWBs are the exception, not the rule \citep{san06,naz09,naz11b}.
In addition, even in the rare cases of X-ray bright CWBs, the X-ray 
emission may not be as bright and hard throughout the whole orbit \citep[for a review, see][]{gud09}.
Therefore, while a bright, hard and phase-locked X-ray emission may be a 
strong indication of binarity, the contrary is certainly not true. The 
presence of a companion cannot always be inferred from X-ray studies, and 
the single status of the other (c)LBVs of our study cannot be definitively
established on the sole basis of their apparent lack of bright X-rays
at the time of the observation(s).

Our study provides additional information allowing to constrain the
presence of an early OB companion as such stars are not X-ray dark.
They display typical X-ray luminosities of $10^{31}$ to $10^{33}$\,\ergs\
(\citealt{har79,ku79}, see also recent statistical studies of \citealt{san06,ant08,naz09,naz11b}). Even if bright ($>10^{33}$\,\ergs) wind-wind 
collisions do not exist in massive systems, the intrinsic X-ray emission
of the companion may show up, as the luminosity ratio is less favorable
for the LBV at these high energies. 
Unfortunately, many of our luminosity limits are quite high, hence not 
always able to test this scenario. Only in 5 cases (P\,Cyg, 
AG\,Car, HD\,160529, HD\,316825, and Sher 25) are the data sensitive enough 
to probe the range of X-ray luminosities of OB stars, taking into
account the interstellar absorption of our targets. 

If these LBVs possess OB companions, there are basically two possibilities: either
(1) the companion is quite close to the LBV or (2) the companion is
relatively distant from the LBV. In the first case, the intrinsic
emission of the companion may be hidden by the strong absorption of the dense wind, 
but then the conditions are favorable to get an X-ray bright 
wind-wind collision. In the second case, no emission from a wind-wind
collision is expected due to wind dilution, but the intrinsic emission 
of the companion would be easily detectable, as the tenuous (at these 
distances) wind of the LBV cannot hide it anymore. 

To evaluate the limit between these two scenarios, we focus on the 
most constrained case, that of P\,Cyg. The wind density $\rho(r)$
can be written as $\dot M / 4 \pi\, r^2 v(r)$ where the velocity is assumed
to follow a beta-law with $\beta=1$, i.e. $v(r)=v_{\infty} 
(1-(R_*/r))$. In this case, the equivalent absorbing column for 
hydrogen, from a position $R_{\rm start}$ in the wind can be written as:
\begin{eqnarray*}
N_w(r)=\frac{X_{\rm H}}{m_{\rm H}} \int^{\infty}_{R_{start}} \rho(r) dr
= \frac{X_{\rm H}\, \dot M}{4\pi\,R_*\,v_{\infty}\, m_H} \, \left[ -\ln(1-\,\frac{R_*}{R_{\rm start}}) \right] . \end{eqnarray*}
Using atmosphere models, \citet{naj97} and \citet{naj01} have calculated 
the stellar parameters of P\,Cyg: $R_*=76\,R_{\odot}$, 
$\dot M\sim2 \times10^{-5}$\,$M_{\odot}$\,yr$^{-1}$, 
$n_{\rm He}/n_{\rm H}=0.3$\footnote{which corresponds to an abundance 
fraction (in weight) $X_{\rm H}$ of $\sim$0.45} and $v_{\infty}=185$\,\kms. 
With these parameters and $R_{\rm start}$ of 2, 5, 10, 50, and 100 $R_*$, 
we find absorbing columns of 20, 6, 3, 0.6, and 0.3 
$\times 10^{22}$\,cm$^{-2}$, respectively. An optically-thin plasma
with a temperature of 0.6\,keV and the strength of a late O-early B star 
(i.e. 10$^{31}$\,\ergs at a distance of 1.7\,kpc), which would suffer 
both the interstellar absorption of P\,Cyg and an absorption by such 
columns (modelled within Xspec using the $vphabs$ model and a 3 times solar 
abundance for helium) would have been detected as long as $R_{\rm start}>25R_*$.
This means that only very close binaries would escape detection, 
and only in the worst cases of orientation (i.e. O-star seen through the 
LBV wind, not through the O-star wind). We therefore consider as highly 
probable that P\,Cyg is not a LBV+OB binary.

In summary, we have information on the distance, reddening and 
X-ray luminosity of 24 objects in our survey (i.e. all studied cases
except the Wachter et al. and Gvaramadze et al. candidates). Of these,
4 appear as probable binaries and 5 as probably single stars.
The status of the remaining objects is unclear: we thus have between 
4 and 19 binaries and between 5 and 20 single stars. Assuming
a binary fraction of 0.5, as is often suggested for
massive stars \citep{san11}, we modelize the
LBV population result by a random picking of stars using a binomial 
distribution, so that we find the probability to get the different plausible 
scenarios (i.e. 4 binaries + 20 single, 5 binaries + 19 single, ... 19 binaries + 5 single).
The probability to observe our situation is found to be 99.91\%, for
the assumed binary fraction of 0.5. Only binary fractions below 0.26 
or larger than 0.69 would lead to a different situation in 10\% of the
cases. However, such very low or very high binary fractions seem unlikely for 
populations of massive objects \citep{san11}. 
Considering only the confirmed LBVs yield similar results, though the 
sample is smaller and the statistics therefore less good. In what 
concerns the binary fraction, we thus did not find any evidence that LBVs 
are different from other populations of massive objects.

It must be noted, however, that binarity was proposed to play a role
in the occurrence of giant eruptions. More specifically, these events
would be triggered by close interactions between the components during 
some periastron passages. An example would be the 19th-century eruptions of
eta\,Car, which occurred within weeks of periastron \citep{kas10,smi11c},
and the 20th-century eruptions of HD\,5980 \citep{koe04,koe10}.
The last case of historical giant eruptions is P\,Cyg, and the impact
of a companion has also been strongly advocated for this star by 
\citet{kas10b}. While our observations most probably rule out the 
presence of a massive companion, we cannot exclude that a lower-mass 
object lurks in the neighbourhood of P\,Cyg. For \citet{kas10b},
a late-B companion with 3--6 $M_{\odot}$ is sufficient to trigger
massive ejections. Such an object would produce no X-ray bright 
colliding wind emission (since its wind is negligible) nor a large 
amount of intrinsic X-rays: it would thus remain undetected 
even in our sensitive observations! The question therefore remains open as to 
whether binarity should be considered as a universal trigger of LBV eruptions 
\citep[see also the discussion of alternative triggers in][]{smi11c}.

\section{Conclusion}
We have performed the first survey of Galactic LBVs and candidate LBVs in the X-ray domain,
using a dedicated \xmm\ dataset as well as archival \ch\ and \xmm\ data.
About half of the known (c)LBVs (31 out of 67) or two-third of the (c)LBVs
listed by \citet[22 out of 35]{cla05} have now been studied in the X-ray domain.

The X-ray emission of six objects is detected, but the association
with the (c)LBVs is doubtful for two of them (GCIRS 33SE and 34W).
We are thus left with four (c)LBVs with detections in the X-ray range:
the well known eta Car and Cyg\,OB2\,\#12, plus GAL 026.47+00.02 
and W243. As it is bright and hard, GAL 026.47+00.02 appears rather 
similar to the two ``historical'' detections (eta Car and Cyg\,OB2\,\#12). 
These 3 cases are well
explained, at least quantitatively, by wind-wind collisions in binary
systems. In contrast, the X-ray emission of W243 appears more modest 
and it is compatible with the typical emission level of OB stars:
since the wind velocity of W243 is low, the observed emission may be that
of an otherwise hidden OB companion.

Excluding these objects, only upper limits were found, with values 
down to $L_{\rm X}=8\times10^{29}$\,\ergs\ and $\log[L_{\rm X}/L_{\rm BOL}]-9.4$ for P\,Cyg.
The constraints on the X-ray emission are in fact particularly strong 
in 5 cases (P\,Cyg, AG\,Car, HD\,160529, HD\,316825, and Sher 25) where even the 
usual X-ray emission of a potential OB companion can apparently be 
excluded, indicating that these stars are probably single and that
the intrinsic level of X-ray emission is very low in such LBVs.

Our results (4 detections and 5 strong limits) are fully compatible 
with binary fractions between 0.26 and 0.69 in LBVs, which agrees well with
what is found for O and WR stars. To the limits of our dataset, the multiplicity 
properties of LBVs do not present significant deviations from other populations
of massive stars. Binarity may however have an impact on the occurrence
of giant eruptions, but additional data are needed in order to fully
test this scenario.

\begin{acknowledgements}
The authors thank very much L. Townsley, P. Broos, and L. Bassani for having accepted to share their proprietary data. They acknowledge support from the Fonds National de la Recherche Scientifique (Belgium), the Communaut\'e Fran\c caise de Belgique, the PRODEX XMM-Integral and Herschel contracts, and the `Action de Recherche Concert\'ee' (CFWB-Acad\'emie Wallonie Europe). YN also thanks the \xmm\ helpdesk for interesting discussions on mosaicking data. ADS and CDS were used for preparing this document.
\end{acknowledgements}

\end{document}